\documentclass[aps, prl, twocolumn, groupedaddress,
  superscriptaddress, shortbibliography, notitlepage]{revtex4-1}
\usepackage[latin1]{inputenc}
\usepackage{amsmath,esint}
\usepackage{amsfonts}
\usepackage{amssymb}
\usepackage{color}
\usepackage{graphicx, graphics}
\usepackage{epsfig}
\usepackage{wasysym}
\usepackage{xr}
\usepackage{hyperref}
\hypersetup{colorlinks=true, urlcolor=blue,
  linkcolor=blue, citecolor=blue}
\usepackage{subcaption}
\newcommand{\vect}[1]{\mathbf{#1}}

\begin{document}
\title{Monte Carlo sampling for stochastic weight functions}

\author{Daan Frenkel}
\email{df246@cam.ac.uk}
\affiliation{Department of Chemistry, University of Cambridge,
  Lensfield Road, Cambridge, CB2 1EW, UK}
\author{K. Julian Schrenk}
\affiliation{Department of Chemistry, University of Cambridge,
  Lensfield Road, Cambridge, CB2 1EW, UK}
\author{Stefano Martiniani}
\affiliation{Department of Chemistry, University of Cambridge,
  Lensfield Road, Cambridge, CB2 1EW, UK}

\begin{abstract}
Conventional Monte Carlo simulations are stochastic in the sense that the acceptance of a trial move is decided by comparing a computed acceptance probability with a random number, uniformly distributed between 0 and 1. Here we consider the case that the weight determining the acceptance probability itself is fluctuating. This situation is common in many numerical studies. We show that it is possible to construct a rigorous Monte Carlo algorithm that visits points in state space with a probability proportional to their average weight. The same approach has the potential to transform the methodology of a certain class of high-throughput experiments or the analysis of noisy datasets.
\end{abstract}

\maketitle

Dynamic Monte Carlo simulations aim to sample the states of the system under study such that the frequency with which a given state is visited is proportional to the weight (often `Boltzmann' weight) of that state.  The equilibrium distribution of a system, i.e. the distribution for which every state occurs with a probability proportional to its (Boltzmann) weight, is invariant under application of single Monte Carlo step. Algorithms that satisfy this criterion are said to satisfy `balance'~\cite{Deem}. Usually, we impose a stronger condition: `detailed balance', which implies that the average rate at which the system makes a transition from an arbitrary `old' state ($o$) to a `new' state ($n$) is exactly balanced by the average rate for the reverse rate.
The detailed balance condition is a very useful tool to construct valid Markov Chain Monte Carlo (MCMC) algorithms. We can write the detailed balance condition as follows;
\begin{equation}
\begin{split}
P(\vect{x}_o) P_{\text{gen}}(o\rightarrow n) P_{\text{acc}}(o\rightarrow n) \\ =  
P(\vect{x}_n) P_{\text{gen}}(n\rightarrow o) P_{\text{acc}}(n\rightarrow o) 
\end{split}
\end{equation}
where $P(\vect{x}_i)$ denotes the equilibrium probability that the system is in state $i$ (in this case, $i$ can stand for $o$ or $n$) characterised by a (usually high-dimensional) coordinate  $\vect{x}_i)$. $P_{\text{gen}}(i\rightarrow j)$ denotes the probability to generate a trial move from state $i$ to state $j$. In the simplest case, this may be the probability to generate a random displacement that will move the system from $\vect{x}_i$ to $\vect{x}_j$, but in general the probability to generate a trial move may be much more complex (see e.g. Ref.~\cite{FrenkelSmit}). Finally $P_{\text{acc}}(i\rightarrow j) $ denotes the probability that a trial move from state $i$ to state $j$ will be accepted. 

Many  simple MC algorithms satisfy in addition microscopic reversibility, which means that $P_{\text{gen}}(i\rightarrow j)=P_{\text{gen}}(j\rightarrow i)$. In that case, detailed balance implies that
\begin{equation}
\frac{P_{\text{acc}}(o\rightarrow n)}{P_{\text{acc}}(n\rightarrow o)}= \frac{P(\vect{x}_n)}{P(\vect{x}_o)}
\end{equation}
There are many acceptance rules that satisfy this criterion. The most familiar one is the so-called Metropolis rule~\cite{Metropolis}:
\begin{equation}
P_{\text{acc}}(o\rightarrow n) =\mbox{Min}\left\{1,\frac{P(\vect{x}_n)}{P(\vect{x}_o)} \right\}
\end{equation}
The acceptance for the reverse move follows by permuting $o$ and $n$.  In the specific case of Boltzmann sampling of configuration space, where the equilibrium distribution is proportional to  the Boltzmann factor $P(\vect{x}_i) \sim \exp(-U_i/k_\text{B}T)$, where $U_i$ is the potential energy of the system in the state characterised by the coordinate  $\vect{x}_i$, $T$ is the absolute temperature and $k_\text{B}$ is the Boltzmann constant. In that case, we obtain the familiar result
\begin{equation}
P_{\text{acc}}(o\rightarrow n) =\mbox{Min}\left\{1,\exp[-(U_n-U_o)/k_\text{B}T] \right\}
\end{equation}

\section*{Monte Carlo simulations with `noisy' acceptance rules.}
There are many situations where conventional MCMC cannot be used because the quantity that determines the weight of a state $i$ is, itself, the average of a fluctuating quantity. Specifically, we consider the case of weight functions fluctuating according to a Bernoulli process, i.e. in an intermittent manner, although our approach is not limited to Bernoulli processes. Examples that we consider are `committor' functions, or the outcome of a stochastic minimisation procedure. 

Equally interesting are examples where a MCMC algorithm would be employed to steer  a (high throughput) experiment where we aim to optimise an output (e.g. crystal nucleation) that is only determined in a probabilistic sense by the initial conditions (typically specified by a large number of parameters). Yet another example would be an experiment that aims to find optimal solutions based on stochastic outcomes  (e.g. finding the biologically most functional and/or least harmful composition of a multi-drug cocktail). Problems of this nature -- and there are many of them -- are, at present not tackled using MCMC sampling. Yet, there is no doubt MCMC sampling is the method of choice to explore high-dimensional parameter space.

Note that the problem that we are discussing here is different from the case considered by Ceperley and Dewing (CD)~\cite{CeperleyDewing1999}. CD analysed the problem of performing MCMC sampling of Boltzmann weights in cases where the energy function is noisy. We come back to this point later: suffice it to say that in the case studied in ref.~\onlinecite{CeperleyDewing1999}, the crucial point is that the Boltzmann weight is a nonlinear function of the energy and that therefore the Boltzmann factor corresponding to the average energy is not the same as the average of the Boltzmann factor obtained by sampling over energy fluctuations. Ref.~\onlinecite{CeperleyDewing1999} showed how to construct an approximate algorithm for such cases, which becomes exact if the fluctuations are normally distributed. Here we consider the case where the probability to sample a point is given rigorously by the average of the stochastic estimator of the weight function. 

To give a specific example, we consider the problem of computing the volume of the basin of attraction of a particular energy minimum $i$ in a high-dimensional energy landscape~\cite{Xu,Asenjo,Martiniani16a, Martiniani16b}.
The algorithms developed in Refs.~\onlinecite{Xu,Asenjo,Martiniani16a, Martiniani16b} rely on the fact that, for every point $\vect{x}$ in configuration space, we can determine unambiguously whether or not it belongs to the basin of attraction of minimum $i$: if a (steepest-descent or similar) trajectory that start at point $\vect{x}$ ends in minimum $i$, the `oracle function' $\mathcal{O}_i(\vect{x})=1$, and otherwise it is zero.

However, many minimizers are not deterministic -- and hence the oracle function is probabilistic. (In fact, historical evidence suggests that ancient oracles were probabilistic at best).  In that case, if we start a number of minimisations at point $\vect{x}$, some will have $\mathcal{O}_i(\vect{x})=1$ and others have $\mathcal{O}_i(\vect{x})=0$.  We denote with $P_\mathcal{O}^{(i)}(\vect{x})$ the average value of the Bernoulli process defined by the oracle function $\mathcal{O}_i(\vect{x})$. In words: $P_\mathcal{O}^{(i)}(\vect{x})$ is the probability that the oracle function associated with point $\vect{x}$ has a value of one. 
 
We now redefine the basin volume (probability mass) of minimum $i$ as
\begin{equation}
\label{eq:volume_integral}
v_i\equiv \int d\vect{x} \;P_\mathcal{O}^{(i)}(\vect{x})
\end{equation}
where $\vect{x} $ denotes the coordinate in $d$-dimensional space. Clearly,
\begin{equation}
\sum_{i=1}^\Omega v_i = V_{total}
\end{equation}
where $\Omega$ is the number of distinct minima. This equation expresses the fact that every trajectory must end up somewhere. Hence, we now have an algorithm that allows us to define basin `clouds' rather than basin volumes, but for the rest the language stays the same. 

\subsection{Naive MC algorithm}
If we consider a large number of trial moves form point $\vect{x}$ to point $\vect{x}'$, the average acceptance probability is  $P_\mathcal{O}(\vect{x}')$. If we consider a large number of trial moves in the reverse direction, the acceptance probability is  $P_\mathcal{O}(\vect{x})$. In steady state, the populations should be such that detailed balance holds. If we denote the `density' of sampled points by $\rho(\vect{x})$, then
\begin{equation}
 \rho(\vect{x})P_{\text{acc}}(\vect{x}\rightarrow\vect{x}')=
  \rho(\vect{x}')P_{\text{acc}}(\vect{x}'\rightarrow\vect{x})
  \end{equation}
  Hence if we choose the acceptance probability to be equal to the (instantaneous) value of the oracle function in the trial state, then
  \begin{equation}
 \rho(\vect{x})P_\mathcal{O}(\vect{x}')=
  \rho(\vect{x}')P_\mathcal{O}(\vect{x})
  \end{equation}
or
\begin{equation}
 {\rho(\vect{x})\over   \rho(\vect{x}')} = 
 {P_\mathcal{O}(\vect{x})\over P_\mathcal{O}(\vect{x}')}
\end{equation}
In words: points are sampled with a probability proportional to the value of the oracle function. Note that in this naive version of the algorithm, the acceptance rule is {\em not} the Metropolis rule that considers the ratio of two weights. Here it is the probability itself. Hence, whenever the probability becomes very low, the acceptance of moves decreases proportionally.

There is another class of problems that can be sampled with this algorithm: those that are deterministic but for which the domain where the oracle function is one is highly non-compact. In this case, the key requirement is that the sampling algorithm is ergodic: it should avoid getting stuck in small islands where the oracle function is one. If that can be achieved, then we can use exactly the same approach as before, be it that now the oracle function behaves like a more or less random telegraph function in space. Still, we can define $
v_i\equiv \int d\vect{x} \;P_\mathcal{O}(\vect{x})$ as before. 

\subsection{Configurational bias approach}

In the way it has been formulated above, there is a problem with this approach: as the system moves into a region where $P_{\mathcal{O}}(\vect{x})$ is very low, the acceptance of moves  becomes very small and hence the `diffusion coefficient' that determines the rate at which configuration space is sampled, would become small. As a consequence, sampling of the wings of the distribution may not converge. 

One way to mitigate the sampling problem is to use an approach that resembles configurational bias MC (CBMC) \cite{Frenkel92}, but is different in some respects.  The key point to note is that, if we know all random numbers that determine the value of the oracle function -- including the random numbers that control the behaviour of the stochastic minimiser -- then in the extended space of coordinates and random numbers, the value of the oracle function is always the same for a given point.

One way to exploit this would be to generate a random walk between points that are surrounded by a `cloud' of $k$ points where we compute the oracle function ($k$ is arbitrary, but as we shall see later, it may pay to make it large). We denote the central point (i.e. the one to which or from which moves are attempted) by $\vect{x}_\text{B}$, where `B' stands for `backbone'. The reason for calling this point a `backbone' point is that we will be sampling the $k$ points connected to it, but we will not compute the oracle function at this very point. Hence, $\vect{x}_\text{B}$ may even be located in a region where the oracle function is strictly zero. The coordinates of the $k$ cloud points around $\vect{x}_\text{B}$ are given by:
\begin{equation}
\vect{x}_{\text{B},i}= \vect{x}_\text{B} +{\bf \Delta}_i
\end{equation}
with $i=\{1,2, \cdots,k\}$. The vectors ${\bf \Delta}$ are generated by some stochastic protocol: e.g. the vectors may be uniformly distributed in a hypersphere with radius $R_h$. The precise choice of the protocol does not matter, as long as the rules are not changed during the simulation. For a fixed protocol, the set $\vect{x}_{\text{B},i}$ is uniquely determined by a set of random numbers $\mathcal{R}_\text{B}$. It is convenient (but not essential) to choose the protocol such that any acceptable trial direction about a backbone point is equally likely to be generated. Finally, we note that the value of the oracle function $\mathcal{O}_{i}$ for a given point $\vect{x}_{\text{B},i}$ is uniquely determined by another set of random numbers $\mathcal{R}_\mathcal{O}$. 

We now define an extended state space 
\begin{equation}
{\bf\tilde{\vect{x}}}_\text{B}\equiv \{\vect{x}_\text{B} , \mathcal{R}_\text{B}, \mathcal{R}_\mathcal{O}\} \;.
\end{equation}
In this space, the oracle functions are no longer fluctuating quantities.

We can now construct a MCMC to visit (but not sample) backbone points. To this end, we compute the `Rosenbluth weight' of point ${\bf\tilde{\vect{x}}}_\text{B}$ as
\begin{equation}
\label{eq:rosenbluth}
W ({\bf\tilde{\vect{x}}}_\text{B})= \sum_{i=1}^k \mathcal{O}_i \omega_i,
\end{equation}
where $\mathcal{O}_i \equiv \mathcal{O}({\bf\tilde{\vect{x}}}_{\text{B},i})$ and $\omega_i \equiv \omega ({\bf\tilde{\vect{x}}}_{\text{B},i})$ is some arbitrary (Boltzmann) bias.

We can then construct a MCMC algorithm where the acceptance of a trial move from the `old' 
${\bf\tilde{\vect{x}}}^{(o)}_\text{B}$ to the `new' ${\bf\tilde{\vect{x}}}^{(n)}_\text{B}$ is given by 
\begin{equation}\label{eq:acceptance}
P_{\text{acc}}(o\rightarrow n)= \mbox{Min}\left\{1, {W ({\bf\tilde{\vect{x}}}^{(n)}_\text{B})\over W ({\bf\tilde{\vect{x}}}^{(o)}_\text{B})}\right\}
\end{equation}

As the probabilities to generate the trial directions for forward and backward moves, and the generation of random numbers that determine the value of the oracle function are also uniform, the resulting MC algorithm satisfies super-detailed balance and a given backbone point ${\bf\tilde{\vect{x}}}_\text{B}$ will be visited with a probability proportional to $W ({\bf\tilde{\vect{x}}}_\text{B})$.

Note that during a trial move, the state of the old point is not changed, hence it retains the same trial directions and the same set $\{\mathcal{R}_\mathcal{O} \}$. If the trial move is rejected, it is this `extended point' that is sampled again.

\subsection{Sampling}

\begin{figure}[t]
\centering
 \includegraphics[width=\linewidth]{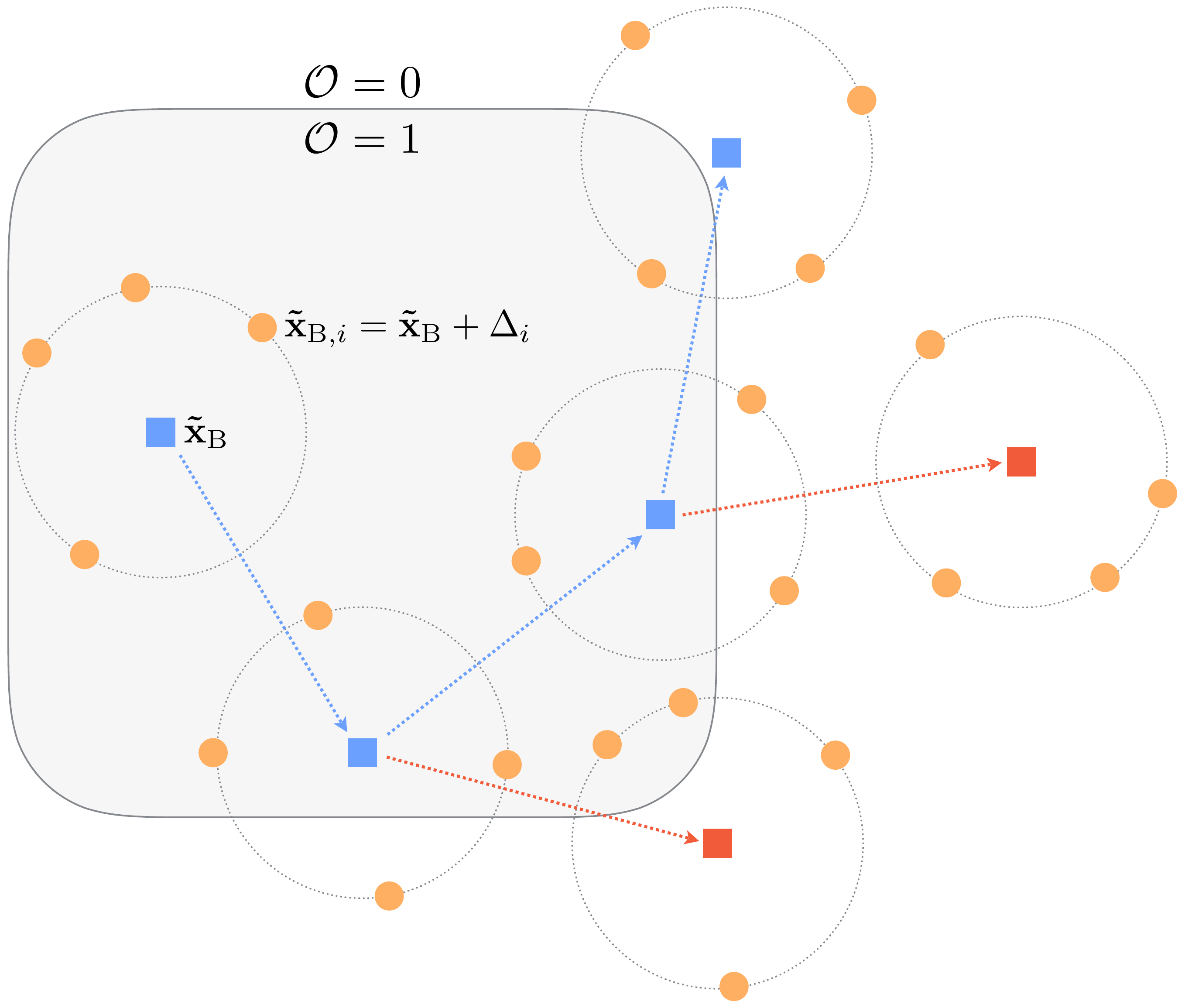}
    \caption{\label{fig::cloud_sampling} `Cloud' sampling: illustration of the configurational bias approach for a simple oracle defined by the gray shaded region, such that $\mathcal{O}=1$ inside the gray boundary and $\mathcal{O}=0$ outside.  Blue and red squares are the accepted and rejected backbone points $\tilde{\vect{x}}_{\mathrm{B}}$, respectively. The `cloud' points $\tilde{\vect{x}}_{\mathrm{B},i}=\tilde{\vect{x}}_{\mathrm{B}} + \Delta_i$ are represented by orange circles. In this example we randomly sample $k=4$ `cloud' points from a circle of fixed radius centred on the backbone point (dotted circles). Each `cloud' is sampled with probability proportional to the Rosenbluth weight defined in Eqn.~\ref{eq:rosenbluth}. Note that valid backbone points are not required to fall in the region where $\mathcal{O}=1$ since the Rosenbluth weight does not depend on the value of the oracle at the backbone point. }
\end{figure}

To discuss sampling, it is best to first discuss a `thought-algorithm' {\em i.e.} a valid algorithm that we can construct, but that we would never use in practice. In our thought algorithm, we consider the transition between one particular point, say $i_o$ in the cloud around the old backbone position and another point $i_n$ in the cloud around the new backbone position. Note that the statistical weight of these points depends on ${\bf\tilde{\vect{x}}}_\text{B}^{(o)}$ and ${\bf\tilde{\vect{x}}}_\text{B}^{(n)}$, respectively. We denote these statistical weights by 
$P({\bf\tilde{\vect{x}}}_\text{B}^{(o)})$ and $P({\bf\tilde{\vect{x}}}_\text{B}^{(n)})$. We can now write down the detailed balance condition:
\begin{eqnarray}
P({\bf\tilde{\vect{x}}}_\text{B}^{(o)}) P_{\text{gen}}({\bf\tilde{\vect{x}}}_\text{B}^{(n)}) P_{\text{sel}}(i_n)P_{\text{acc}}(o\rightarrow n)\nonumber\\
=P({\bf\tilde{\vect{x}}}_\text{B}^{(n)}) P_{\text{gen}}({\bf\tilde{\vect{x}}}_\text{B}^{(o)}) P_{\text{sel}}(i_o)P_{\text{acc}}(n\rightarrow o)\; ,
\end{eqnarray}
where $P_{\text{sel}}(i_n)$ denotes the probability to select point $i_n$ from among the cloud of points around $\vect{x}^{(n)}_\text{B}$ (and similarly, for $P_{\text{sel}}(i_o)$). We now make the following choice for $P_{\text{sel}}$:
\begin{equation}
P_{\text{sel}}(i_n)=\frac{\mathcal{O}(i_n) \omega(i_n)}{ \sum_{i'=1}^k \mathcal{O}(i'_n) \omega(i'_n)} = \frac{\mathcal{O}(i_n) \omega(i_n)}{W({\bf\tilde{\vect{x}}}^{(n)}_\text{B})}\; ,
\end{equation}
With this definition of the selection probability, we can write:
\begin{equation}\label{eq:DetBal}
\frac{P({\bf\tilde{\vect{x}}}_\text{B}^{(o)}) }{P({\bf\tilde{\vect{x}}}_\text{B}^{(n)}) }=\frac{\mathcal{O}(i_n) \omega(i_n)}{\mathcal{O}(i_o) \omega(i_o)}\;
\end{equation}
where we have used the fact that the generation probabilities for forward and backward moves are equal and we have inserted Eqn.~\ref{eq:acceptance} for the ratio of the acceptance probabilities. Eqn.~\ref{eq:DetBal} implies that in equilibrium, the probability to occupy state $i_o$ is proportional to $\mathcal{O}(i_o)\omega(i_o)$, where it should be stressed that the value of the oracle function depends on both the spatial coordinates of point $i_o$ and  on the set of random numbers $\{\mathcal{R}_\mathcal{O} \}$ that, together,  determine the value of $\mathcal{O}(i_o)$. If we were to average over all possible values of the random numbers $\{\mathcal{R}_\mathcal{O} \}$ then it is clear that the probability to sample a state with the spatial coordinates of the point $i_o$ is proportional to $\left\langle\mathcal{O}(i_o)\right\rangle\omega(i_o)$. In other words, the algorithm described above samples all points in configuration space with a probability proportional to the local average of the oracle function and to the (usually Boltzmann) bias evaluated at that point. 

Whilst the above description of the sampling strategy allows us to establish that all points in space are sampled with the correct frequency, it is not an efficient algorithm. The reason is obvious: in order to compute the weights $W$, the oracle function must be computed for $k$  points, and yet in the naive version of the algorithm, only one point is sampled. In practice, we take steps between backbone points sampled according to Eqn.~\ref{eq:acceptance} and keep all $k$ cloud points for all the accepted backbone points, as described below. An illustration of the method is given in Fig.~\ref{fig::cloud_sampling}. Efficiency can be further improved using the approach underlying `waste-recycling' Monte Carlo~\cite{DFWasteRecycling}, we can in fact include all points in the sampling, even if the actual trial backbone move is rejected. 
 
For every backbone point ${\bf\tilde{\vect{x}}}_\text{B}$ visited, we can compute the observable (say $A$) of the set of $k$ cloud points as follows:
\begin{equation}
A_{\text{sampled}} = \frac{ \sum_{i=1}^k \mathcal{O}_i \omega_i A_i}{\sum_{i=1}^k \mathcal{O}_i \omega_i}
\end{equation}
The average of $A$ during a MCMC simulation of $L$ steps is:
 \begin{equation}
 \label{eq:mc_average}
 {1\over L} \sum_{j=1}^L  \left(\frac{ \sum_{i=1}^k \mathcal{O}_i \omega_i A_i}{\sum_{i=1}^k \mathcal{O}_i \omega_i}\right)_j
\end{equation}
where the index $j$ labels the different backbone states visited.

\subsection{Parallel Tempering}

Parallel Tempering (PT) \cite{Lyubartsev92, Marinari92} is a Monte Carlo scheme that targets the slow equilibration of systems characterised by large free energy barriers that prevent the efficient equilibration of a MCMC random walk. In PT, $m$ replicas of the system are simulated simultaneously at different temperatures, different chemical potentials \cite{Yan99} or different Hamiltonians \cite{Bunker00, Fukunushi02}. Configurations are then swapped among replicas, thus making `high temperature' regions available to `low temperature' ones and \emph{vice versa}. In the basin volume calculations of Refs. \onlinecite{Asenjo, Martiniani16a, Martiniani16b, Martiniani16c}, Hamiltonian PT is essential to achieving fast equilibration of the replicas' MCMC random walks performed inside the body of the basin with different applied biases. 

The configurational bias approach to `cloud' sampling embodied by Eqn.~\ref{eq:acceptance} can be easily generalised to PT to find an acceptance rule for the swap of configurations between replicas $i$ and $j$
\begin{equation}
\label{eq:pt_acceptance}
P_{\text{acc}}(i\rightarrow j)= \mbox{Min}\left\{1, {W ({\bf\tilde{\vect{x}}}^{(i)}_\text{B}, \omega^{(j)})W ({\bf\tilde{\vect{x}}}^{(j)}_\text{B}, \omega^{(i)})\over W ({\bf\tilde{\vect{x}}}^{(i)}_\text{B}, \omega^{(i)}) W ({\bf\tilde{\vect{x}}}^{(j)}_\text{B}, \omega^{(j)})}\right\}
\end{equation}
where we defined the Rosenbluth weight $W ({\bf\tilde{\vect{x}}}^{(i)}_\text{B}, \omega^{(j)}) = \sum_{l=1}^k \mathcal{O}({\bf\tilde{\vect{x}}}^{(i)}_{\text{B},l}) \omega^{(j)}({\bf\tilde{\vect{x}}}^{(i)}_{\text{B},l})$. It is important to note that PT is truly an equilibrium Monte Carlo method: the microscopic equilibrium of each ensemble is not disturbed by the swaps.

\subsection{Combine with `Waste-recycling' MC}
Using a CBMC-style approach would increase the speed with which the relevant configuration space is explored. However, it has the drawback that it may be wasteful: generating a trial move involves computing $k$ oracle functions and, in normal CBMC the points this generated would not be sampled at all if the trial move is rejected.

However, we can do better by using `waste-recycling' MC~\cite{DFWasteRecycling}.
In that case we  can combine the information of the accepted and the rejected states in our sampling. Specifically, we denote the probability to accept a move from an old state $o$ to a new state $n$ by $P_{\text{acc}}(o\rightarrow n)$, then, normally  we would sample  $A_{\text{sampled}}(n)$ if the move is accepted and  $A_{\text{sampled}}(o)$ otherwise. However, we can do better by combining the information and sample
\begin{equation}
A_{\text{wr}} = P'_{\text{acc}}(o\rightarrow n) A_{\text{sampled}}(n)+ [1-P'_{\text{acc}}(o\rightarrow n)] A_{\text{sampled}}(o)
\end{equation}
where $P'_{\text{acc}}$ denotes the acceptance probability for {\em any} valid MCMC algorithm (not just Metropolis). In fact, it is convenient to use the symmetric (Barker) rule~\cite{Barker} to compute $P'_{\text{acc}}$. In that case, we would sample
\begin{equation}
A_{wr}= \frac{ \left(\sum_{i=1}^k \mathcal{O}_i \omega_i A_i\right)_{old} +\left(\sum_{i=1}^k \mathcal{O}_i \omega_i A_i\right)_{new} }{\left(\sum_{i=1}^k \mathcal{O}_i\omega_i\right)_{old}+\left(\sum_{i=1}^k \mathcal{O}_i\omega_i\right)_{new}  }
\end{equation}
Hence, all $2k$ points that have been considered are included in the sampling.

\begin{figure}[t]
\centering
 \includegraphics[width=\linewidth]{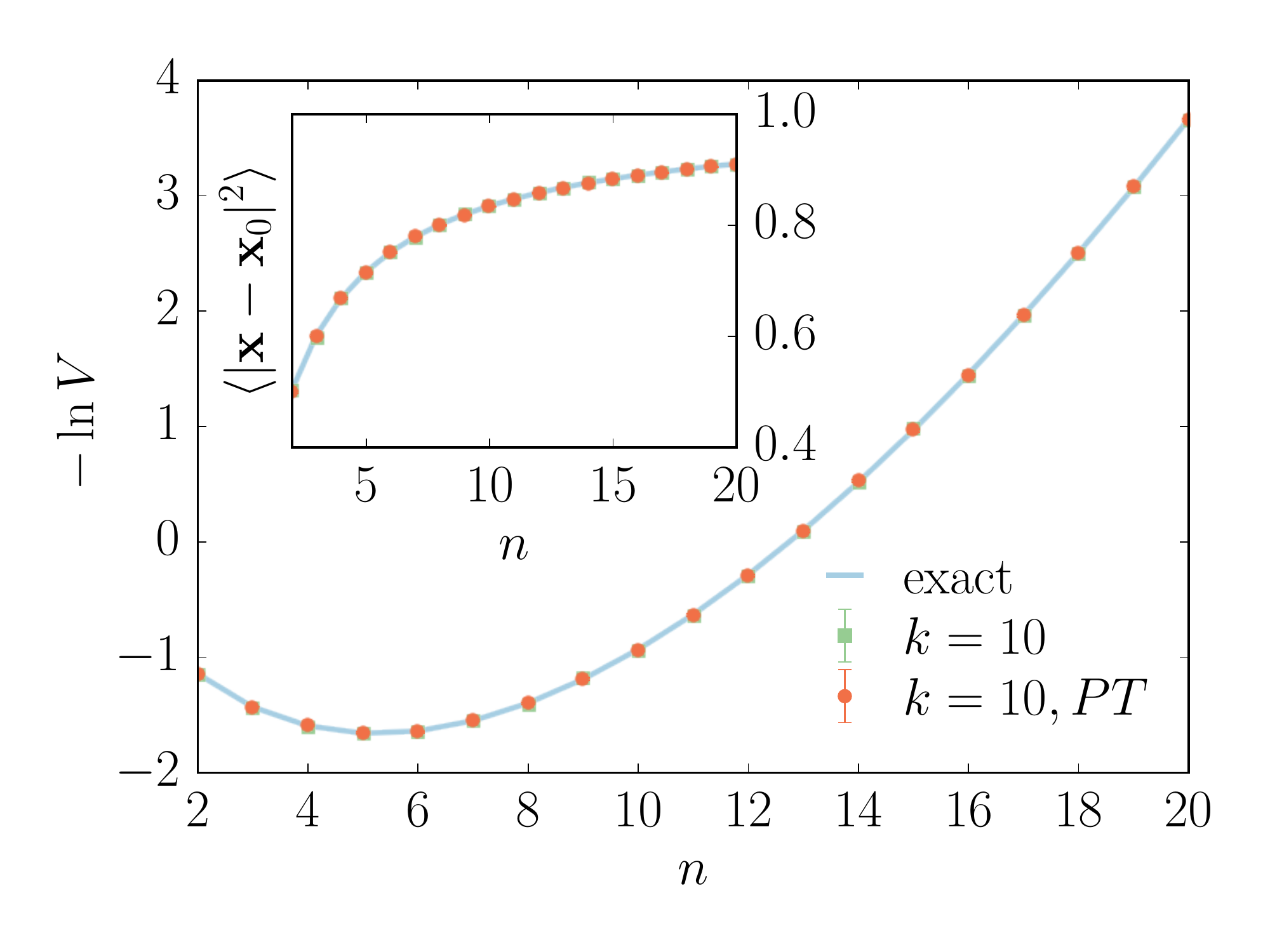}
    \caption{\label{fig::sphere} Deterministic oracle: Volume calculation for an $n$-ball with radius $R=0.5$ and $n\in [2,20]$. Numerical results (symbols) were obtained by the configurational bias approach of Eqn.~\ref{eq:acceptance} and Eqn.~\ref{eq:pt_acceptance} (PT), with $k$ `cloud' points, and MBAR. Inset: mean square displacement computed by Eqn.~\ref{eq:mc_average}. Solid blue lines are analytical results and error bars refer to twice the standard error (as estimated by MBAR for the volume).}
\end{figure}

\begin{figure}[t]
\centering
 \includegraphics[width=\linewidth]{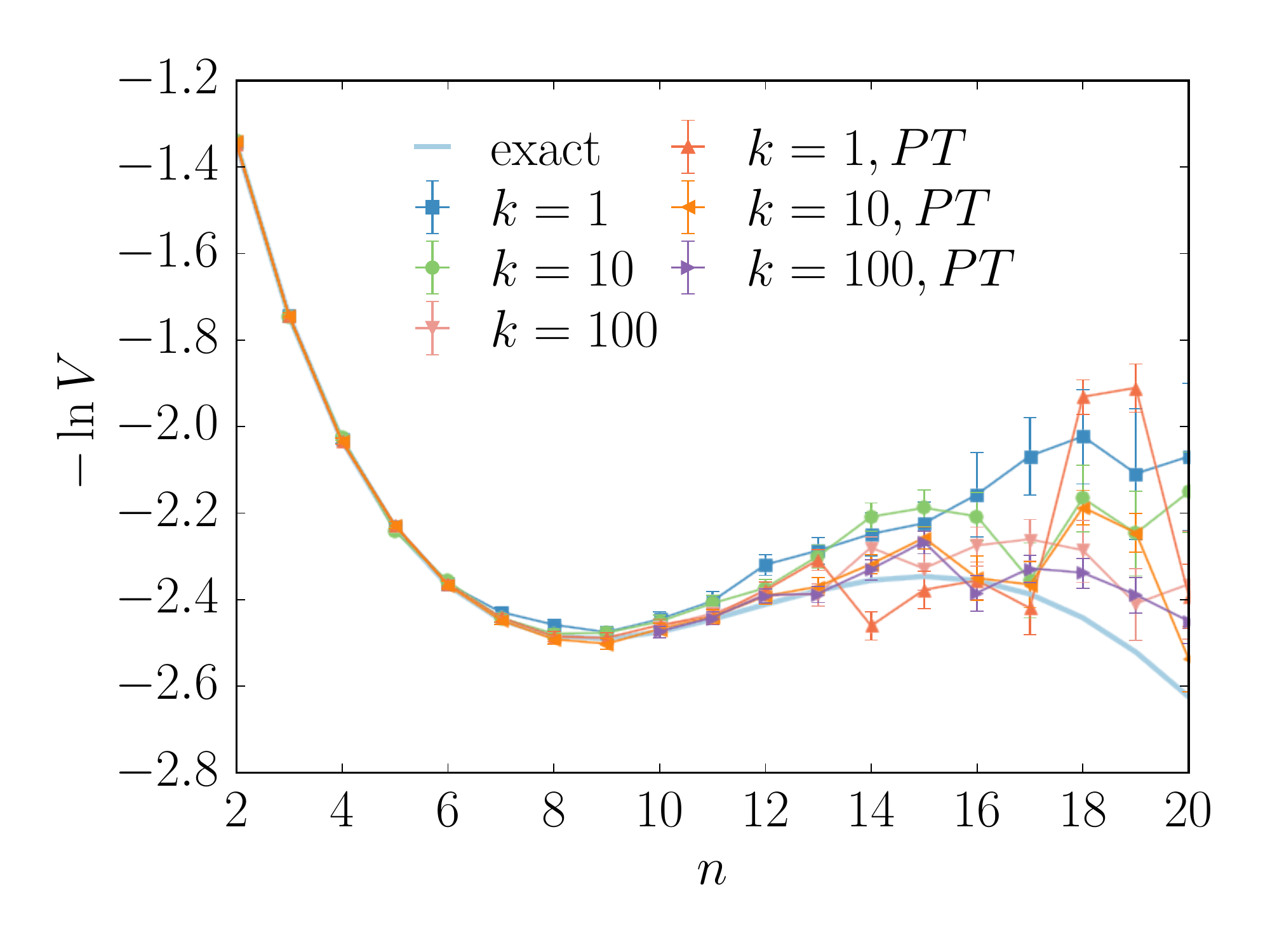}
    \caption{\label{fig::sphere_exp_decay} Stochastic oracle: Volume calculation for the oracle defined in Eqn.~\ref{eq:sphere_exp_decay} with radius $R=0.5$, $\lambda=0.1$ and dimensions $n\in [2,20]$. Symbols (lines are guide to the eye) are numerical results obtained by the configurational bias approach of Eqn.~\ref{eq:acceptance} and Eqn.~\ref{eq:pt_acceptance} (PT), with $k$ `cloud' points, and MBAR. Solid blue line is the analytical result and error bars refer to twice the standard error as estimated by MBAR. At large $n$ accuracy increases by increasing $k$ as the random walker diffuses more efficiently through regions of space where $\langle \mathcal{O} \rangle \ll 1$. Implementing PT also improves equlibration for small $k$ by allowing the walker to escape low density regions when stuck.}
\end{figure}

\begin{figure}[t]
\centering
 \includegraphics[width=\linewidth]{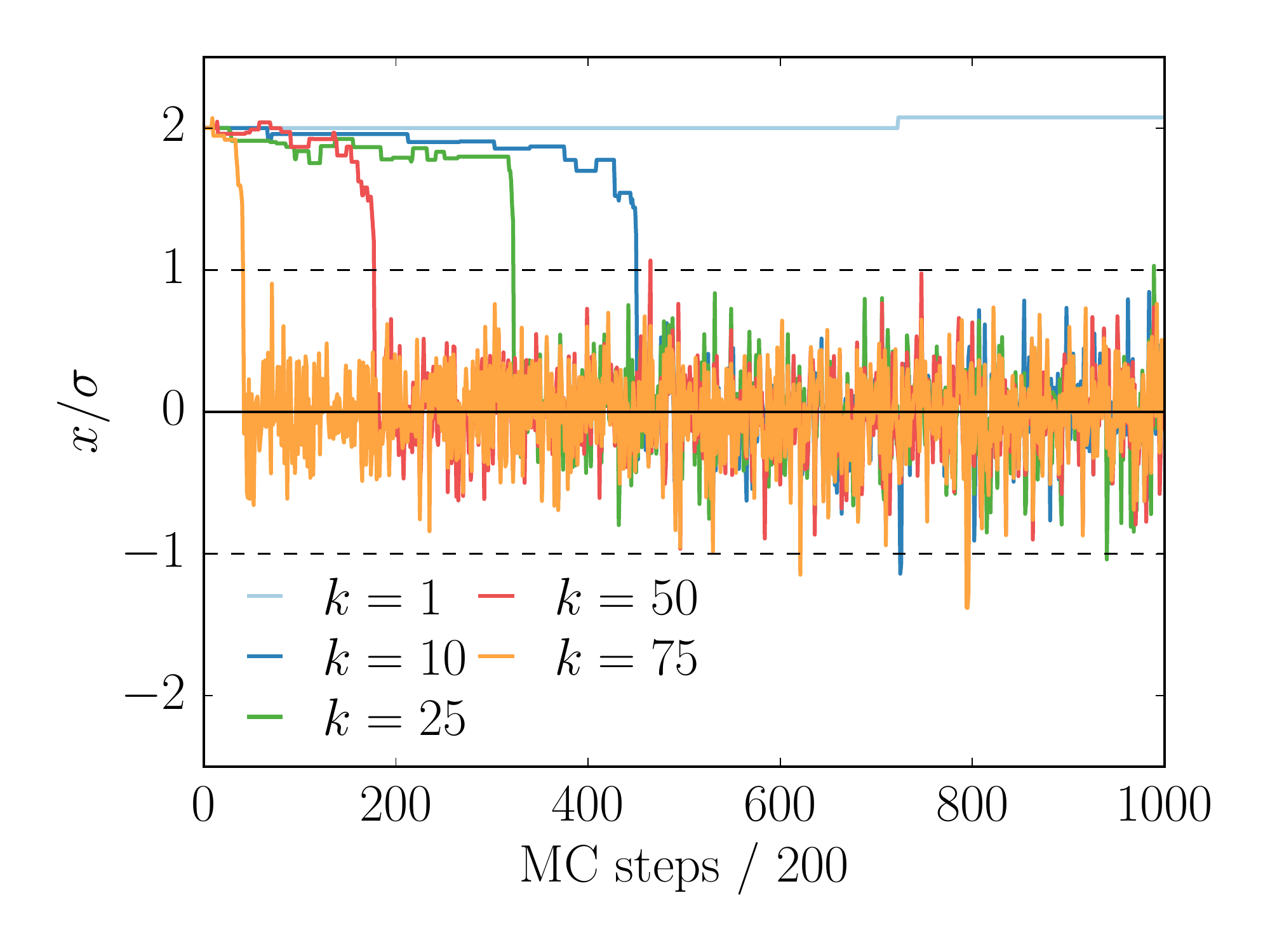}
    \caption{\label{fig::barrier_crossing} Transition state finding: the simple case of one dimensional barrier crossing is defined (symmetrically) by the stochastic oracle in Eqn.~\ref{eq:barrier_crossing}. A series of random walks are performed according to Eqn.~\ref{eq:acceptance} with different number of `cloud' points $k$. The walkers are constrained to reject moves for which the energy is below that of the initial position, thus excluding reactants and products from the sampling. The figure shows the position of the walker backbone along the reaction coordinate as a function of the number of MCMC steps. For increasing $k$ the random walkers diffuse more efficiently and therefore converge faster to the transition state. Traditional single-point sampling does not move at all from the initial condition.}
\end{figure}

\section*{Numerical Results}

\subsection{Basin volume calculations}

We test the proposed configurational bias approach by numerically computing the basin volume (probability mass) for a stochastic oracle function as defined in Eqn.~\ref{eq:volume_integral}. We choose a few simple oracle functions, for which the integral in  Eqn.~\ref{eq:volume_integral} can be solved analytically. 

The volume calculations are performed using the multistate-Bennett acceptance ratio method (MBAR)~\cite{ShirtsChodera} as described in Ref.~\onlinecite{Martiniani16b}. In essence, we compute the dimensionless free energy difference between a region of known volume $\widehat{f}_{\text{ref}} = -\ln V_{\text{ref}} + c$ and the equilibrium distribution of points sampled uniformly within the basin $\widehat{f}_{\text{tot}} = -\ln V_{\text{tot}} + c$, estimated by MBAR up to a multiplicative constant $c$. Since $f_{\text{ref}} = -\ln V_{\text{ref}}$ is known, we obtain the basin volume as $f_{\text{tot}} = f_{\text{ref}} + (\widehat{f}_{\text{tot}}-\widehat{f}_{\text{ref}})$. We use $15$ replicas with positive coupling constants for all examples discussed herein, see Ref.~\onlinecite{Martiniani16b} for details of the method.

First, we test the method for a deterministic oracle, namely a simple $n$-ball of known volume $V_{n\text{-ball}}=\pi^{n/2}R^n/\Gamma(n/2+1)$ with radius $R=0.5$ and $n\in [2,20]$. As shown in Fig.~\ref{fig::sphere} we correctly recover the volume and the mean square displacement using the acceptance rule defined in Eqn.~\ref{eq:acceptance} for $k=10$ `cloud' points, with and without parallel tempering swap moves, with acceptance rule defined in Eqn.~\ref{eq:pt_acceptance}. Hence, the algorithm is clearly sampling the correct equilibrium distributions.

We test the method for a stochastic oracle function defined as
\begin{equation}
\label{eq:sphere_exp_decay}
\mathcal{O}(\vect{x})= \left\{ 
  \begin{array}{l l}
    1 & \quad \text{if $|\vect{x}| < R$} \\ 
    \text{Uniform}[0,1] < \exp[-(|\vect{x}|-R)/\lambda] & \quad \text{if $|\vect{x}| \geq R$}
  \end{array} \right.\, 
\end{equation}
with volume 
\begin{equation*}
V=2(R^n/n + \lambda^n \exp(R/\lambda)\Gamma(n, R/\lambda))\pi^{n/2}R^n/\Gamma(n/2),
\end{equation*} 
where $\Gamma(a,x)$ is the incomplete gamma function. Results for dimensions $n\in[2,20]$, $R=0.5$ and $\lambda=0.1$ are shown in Fig.~\ref{fig::sphere_exp_decay}. Note that, despite the volume being finite, the basin is unbounded in the sense that the average value of the oracle only tends to zero as
as $|\vect{x}| \to \infty$. As the dimensionality of the basin increases, all of the volume will concentrate away from the centre of mass in regions of space where the oracle has a high probability of returning  $0$. Hence, it becomes more difficult for a random walker to diffuse efficiently as the dimensionality of space increases. We can verify this in Fig.~\ref{fig::sphere_exp_decay}: for $n<6$ results seem to be independent of the number of `cloud' points and of whether PT swaps are implemented. However, growing deviations are observed for increasing $n$ and accuracy increases significantly for growing number of `cloud' points $k$ and with the use of PT, whose non-local moves allow the walker to escape regions of low density (for which $\langle \mathcal{O} \rangle \ll 1$) when stuck.

\subsection{Transition state finding}

In this example we show that our approach can be used to efficiently identify the transition state along a known reaction coordinate.

Note that points in the transition-state ensemble (in the one-dimensional case: just one point)  are characterised by the property that the committor has an average value of 0.5. However, any individual trajectory will either be crossing (``1'') or non-crossing (``0''). Hence, the `signal' is stochastic.  As an illustration,  we consider the (trivial) one-dimensional case of a particle with kinetic energy $K$ sampled according to the 1-dimensional Maxwell Boltzmann distribution, crossing a Gaussian barrier with height $U_{\text{tr}}=30kT$ and variance $\sigma^2=1$ \footnote{We choose as our unit of length $\sigma$, hence in our reduced units $kT = \sigma^2$}. We define the oracle symmetrically such as 
\begin{equation}
\label{eq:barrier_crossing}
\mathcal{O}(x)= \left\{ 
  \begin{array}{l l}
    1 & \quad \text{if $K > U _{\text{tr}} - U(x)$} \\ 
    0 & \quad \text{if $K \leq U _{\text{tr}} - U(x)$}
  \end{array} \right.\, 
\end{equation}
and constrain the walk to reject moves for which the energy is below that of the initial position, such that $\mathcal{O}=0$ if $U(x) < U(x_0)$; we choose $x_0 = 2\sigma$. By thus constraining the sampling, we are excluding the `reactant' and `product' states from our sampling. In Fig.~\ref{fig::barrier_crossing} we show results for backbone step-size $0.25\sigma$, `cloud' radius $0.25\sigma$ and varying number of `cloud' points $k$. One can clearly see that as the number of `cloud' points increases the system diffuses faster towards the transitions state whilst for the traditional single-point sampling the walker does not move at all from the initial position. 

\section*{Relation to earlier work}

In their 1999 paper, Ceperley and Dewing~\cite{CeperleyDewing1999} consider a different situation where normal `Metropolis' sampling fails, namely the case where the calculation of the energy function is subject to statistical errors (with zero mean). In that case, we cannot use the conventional Metropolis rule $P_{\text{acc}}=\mbox{Min}\{1,\exp(-\beta\Delta u)\}$, where $u$ is the instantaneous value of the energy difference, because  what is needed to compute the correct acceptance probability is $\exp(-\beta\langle\Delta u\rangle)$,  but what is sampled is $\langle\exp(-\beta\Delta u)\rangle\ne \exp(-\beta\langle \Delta u \rangle)$. Ceperley and Dewing showed that if the fluctuations in the energy of the individual states, and therefore the fluctuations in $\Delta u$ are normally distributed, and if the variance in energy is the same for all states, then we can still get an algorithm that samples the correct Boltzmann distribution, if we use as acceptance rule
\begin{equation}
P_{\text{acc}}=\mbox{Min}\{1,\exp[-\beta\Delta u -(\beta\sigma)^2/2]\}
\end{equation}
where $\sigma^2=2\sigma_s^2$ , with $\sigma_s$ denoting the variance in the energy of the individual states.
Note that the situation considered in Ref.~\onlinecite{CeperleyDewing1999} is very different from the case that we consider here, as we focus on the situations where the average of the (fluctuating) oracle functions is precisely the weight function that we wish to sample. However, the current approach allows us to rederive the CD result. We note that, as before, we can consider extended states characterised by the spatial coordinates of the system and by the random variables that characterise the noise in the energy function. 
First, we note the average Boltzmann factor of extended state $i$ is
\begin{equation}
\left\langle P_i\right\rangle =\exp[-\beta\langle u \rangle_i] \exp[+(\beta\sigma_s)^2/2]
\end{equation}
and therefore
\begin{equation}
{\left\langle P_n\right\rangle \over \left\langle P_o\right\rangle}=\exp[-\beta\langle \Delta u \rangle]
\end{equation}
Hence, the average Boltzmann factor of any state $i$ is still proportional to the correct Boltzmann weight. However, an MCMC algorithm using the instantaneous Boltzmann weights would not lead  to correct sampling as super-detailed balance yields
\begin{equation}
{P_n(\vect{x}_n)\over P_o(\vect{x}_o)}=\exp[-\beta\Delta u]
\end{equation}
and hence
\begin{equation}
\left\langle{P_n\over P_o}\right\rangle=\exp[-\beta\langle \Delta u \rangle +(\beta\sigma)^2/2]
\end{equation}
which is not equal to
\begin{equation}
{\left\langle P_n\right\rangle \over \left\langle P_o\right\rangle}=\exp[-\beta\langle \Delta u \rangle]
\end{equation}
If, however we would use the CD acceptance rule, we would get
\begin{equation}
\begin{split}
\left\langle{P_n\over P_o}\right\rangle & =\exp[-\beta\langle \Delta u \rangle +(\beta\sigma)^2/2]\times \exp[-(\beta\sigma)^2/2] \\ & = \exp[-\beta\langle \Delta u \rangle]= {\left\langle P_n\right\rangle \over \left\langle P_o\right\rangle}
\end{split}
\end{equation}
Hence, with this rule the states would (on average) be visited with the correct probability.
Note that, as the noise enters non-linearly in the acceptance rule, the CD  algorithm is very different from the one that we derived above. Note also that the present derivation makes it clear that the CD
algorithm can be easily generalised to cases where the noise in the energy is not normally distributed, as long as the distribution of the noise is state-independent. 

\section*{Conclusions and outlook}

Thus far the algorithm described above was presented as a method to perform Monte Carlo sampling in cases where the weight function itself is fluctuating. However, the method might also be used to control certain experiments that study stochastic events (e.g. crystal nucleation, cell death or even the effect of advertising).
Often, the occurrence of the desired event depends on a large number of variables (temperature, pressure, $pH$, concentration of various components) and we would like to select the optimal combination. However, as the desired event itself is stochastic, individual measurements provide little guidance. One might aim to optimise the conditions by accumulating sufficient statistics for individual state points. However, such an approach is expensive. The procedure described in the preceding sections suggests that it may be better to perform experiments in a `cloud' of state points around a backbone point. We could then accept or reject the trial move to a new backbone state using the same rule as in Eqn.~\ref{eq:acceptance}.

\begin{acknowledgements}D. F. acknowledges support by EPSRC Programme Grant EP/I001352/1 and EPSRC grant EP/I000844/1. K. J. S. acknowledges support by the Swiss National Science Foundation (Grant No. P2EZP2-152188 and No. P300P2-161078). S. M. acknowledges financial support from the Gates Cambridge Scholarship.
\end{acknowledgements}


\begin{thebibliography}{1}%
\makeatletter
\providecommand \@ifxundefined [1]{%
 \@ifx{#1\undefined}
}%
\providecommand \@ifnum [1]{%
 \ifnum #1\expandafter \@firstoftwo
 \else \expandafter \@secondoftwo
 \fi
}%
\providecommand \@ifx [1]{%
 \ifx #1\expandafter \@firstoftwo
 \else \expandafter \@secondoftwo
 \fi
}%
\providecommand \natexlab [1]{#1}%
\providecommand \enquote  [1]{``#1''}%
\providecommand \bibnamefont  [1]{#1}%
\providecommand \bibfnamefont [1]{#1}%
\providecommand \citenamefont [1]{#1}%
\providecommand \href@noop [0]{\@secondoftwo}%
\providecommand \href [0]{\begingroup \@sanitize@url \@href}%
\providecommand \@href[1]{\@@startlink{#1}\@@href}%
\providecommand \@@href[1]{\endgroup#1\@@endlink}%
\providecommand \@sanitize@url [0]{\catcode `\\12\catcode `\$12\catcode
  `\&12\catcode `\#12\catcode `\^12\catcode `\_12\catcode `\%12\relax}%
\providecommand \@@startlink[1]{}%
\providecommand \@@endlink[0]{}%
\providecommand \url  [0]{\begingroup\@sanitize@url \@url }%
\providecommand \@url [1]{\endgroup\@href {#1}{\urlprefix }}%
\providecommand \urlprefix  [0]{URL }%
\providecommand \Eprint [0]{\href }%
\providecommand \doibase [0]{http://dx.doi.org/}%
\providecommand \selectlanguage [0]{\@gobble}%
\providecommand \bibinfo  [0]{\@secondoftwo}%
\providecommand \bibfield  [0]{\@secondoftwo}%
\providecommand \translation [1]{[#1]}%
\providecommand \BibitemOpen [0]{}%
\providecommand \bibitemStop [0]{}%
\providecommand \bibitemNoStop [0]{.\EOS\space}%
\providecommand \EOS [0]{\spacefactor3000\relax}%
\providecommand \BibitemShut  [1]{\csname bibitem#1\endcsname}%
\let\auto@bib@innerbib\@empty
\bibitem [{Note1()}]{Note1}%
  \BibitemOpen
  \bibinfo {note} {We choose as our unit of length $\sigma $, hence in our
  reduced units $kT = \sigma ^2$}\BibitemShut {NoStop}%
\end{thebibliography}%


\begin{thebibliography}{100}

\bibitem{Deem} V.I. Manousiouthakis and M. W. Deem,
  J. Chem. Phys. {\bf 110}, 2753--2756 (1999)

\bibitem{FrenkelSmit} D.~Frenkel and  B.~Smit  {\em Understanding
  Molecular Simulations: from Algorithms to Applications}, Academic Press, San
Diego, 2nd edition (2002)

\bibitem{Metropolis} N.~Metropolis, A.~W. Rosenbluth,
  M.~N. Rosenbluth, A. H. Teller, and E. Teller. J. Chem. Phys. {\bf 21} 1087--1092 (1953).

\bibitem{CeperleyDewing1999} D. M. Ceperley and M. Dewing,
  J. Chem. Phys. \textbf{110}, 9812 (1999).

\bibitem{Xu} N.~Xu, D.~Frenkel, and A.~J.~Liu, Phys. Rev. Lett.  {\bf
  106}, 245502 (2011)

\bibitem{Asenjo} D. Asenjo, F. Paillusson, D. Frenkel,
  Phys. Rev. Lett. {\bf 112}, 098002 (2014)

\bibitem{Martiniani16a} S. Martiniani, K. J. Schrenk, J. D. Stevenson,
  D. J. Wales and D. Frenkel, Phys. Rev.  E {\bf 93}, 012906 (2016)

\bibitem{Martiniani16b} S. Martiniani, K. J. Schrenk, J. D. Stevenson,
  D. J. Wales and D. Frenkel, Phys. Rev.  E {\bf 94}, 031301(R) (2016)

\bibitem{Frenkel92} D. Frenkel, G. C. A. M. Mooij, B. Smith,  J. Phys. Condensed Matter {\bf 3}, 3053 (1992)

\bibitem{DFWasteRecycling} D. Frenkel, Proc. Nat. Acad. Sci. USA, {\bf
  101}, 17571--17575 (2004)

\bibitem{Lyubartsev92} A. P. Lyubartsev, A. A. Martsinovski,
  S. V. Shevkunov, P. N. Vorontsov-Velyaminov, J. Chem. Phys. {\bf
    96}, 1776--1783 (1992)

\bibitem{Marinari92} E. Marinari, G. Parisi, Europhys. Lett.  {\bf
  19}, 451 (1992)

\bibitem{Yan99} Q. Yan, J. J. de Pablo, J. Chem. Phys. {\bf 111},
  9509--9516 (1999)

\bibitem{Bunker00} A. Bunker, B. D\"{u}nweg, Phys. Rev. E {\bf 63},
  016701 (2000)

\bibitem{Fukunushi02} H. Fukunishi, O. Watanabe, S. Takada,
  J. Chem. Phys. {\bf 116}, 9058--9067 (2002)

\bibitem{Martiniani16c} S. Martiniani, K. J. Schrenk, K. Ramola, B. Chakraborty and D. Frenkel, preprint arXiv:1610.06328 (2016)

\bibitem{Barker} A.A. Barker, Aust. J. Phys. {\bf 18}, 119 --133
  (1965)

\bibitem{ShirtsChodera} M.R. Shirts and J.D. Chodera,
  J. Chem. Phys. {\bf 129}, 129105 (2008)

\end{thebibliography}
\end{document}